\documentclass[a4paper,12pt]{amsart}

\usepackage{amssymb,amsmath}

\usepackage{mathptmx}       
\usepackage{amssymb,amsmath,graphicx}
\usepackage{enumerate,float,color}
\usepackage[all]{xy}

\theoremstyle{definition}
\newtheorem{thm}{Theorem}[section]
\newtheorem{remark}[thm]{\rm Remark}
\newtheorem{definition}[thm]{Definition}

\newcommand{\R}{\mathbb{R}}

\newcommand{\GL}{\mathrm{GL}}
\newcommand{\Aff}{\mathrm{Aff}}
\newcommand{\SO}{\mathrm{SO}}

\newcommand{\Sym}{\mathrm{Sym}}

\newcommand{\T}{\mathcal{T}}
\newcommand{\F}{\mathcal{F}}
\newcommand{\Blend}{\mathrm{Blend}}

\begin{document}

\title[{Tetrisation of meshes}]{Tetrisation of triangular meshes and its application in shape blending}
\author{Shizuo Kaji}
\address{Yamaguchi University/JST CREST}
\email{skaji@yamaguchi-u.ac.jp}


\maketitle

\begin{abstract}
The As-Rigid-As-Possible (ARAP) shape deformation framework is a versatile
technique for morphing, surface modelling, and mesh editing.
We discuss an improvement of the ARAP framework in a few aspects:
1. Given a triangular mesh in 3D space, 
we introduce a method to associate a tetrahedral structure,
which encodes the geometry of the original mesh.
2. We use a Lie algebra based method to interpolate local transformation, 
which provides better handling of rotation with large angle.  
3. We propose a new error function to compile local transformations into a global piecewise linear map,
which is rotation invariant and easy to minimise.
We implemented a shape blender based on our algorithm 
and its MIT licensed source code is available online.
\end{abstract}

\section{Introduction}

In the seminal paper \cite{ARAP}, they introduced a morphing algorithm
called the As-Rigid-As-Possible (ARAP, for short) shape interpolation.
Since then, the technique has been successfully applied to various shape deformation applications.
In their original paper, tetrahedral volume meshes are used to produce interpolation of shapes.
However, in most computer graphic systems it is common to represent shapes by surface meshes.
To convert a surface mesh to a volume mesh is a non-trivial task (see, for example, \cite{tetgen})
and the resulting volume mesh 
tends to have many extra internal vertices, which makes applications inefficient.
Instead of considering volume meshes, one can ``fatten'' surface meshes.
A common practice is to associate a tetrahedral structure to a triangular surface mesh
by adding the normal vector for every triangle (see, for example, \cite{Sumner2}).
Although this simple trick has been widely used, 
it does not capture important geometric features of the mesh.
For example, the relation between adjacent triangles is neglected.

One of the main purposes of this paper is to introduce a new construction
to associate a tetrahedral structure 
to a triangular mesh, which we call {\em tetrisation} (\S\ref{sec:tetrisation}).
Our method encodes inter-triangular properties
such as the angle between adjacent triangles
so that one can keep track 
of global geometry such as curvature while working locally on tetrahedra.

\begin{figure}[htbp]
 \begin{center}
   \includegraphics[width=0.45\linewidth,keepaspectratio=true]{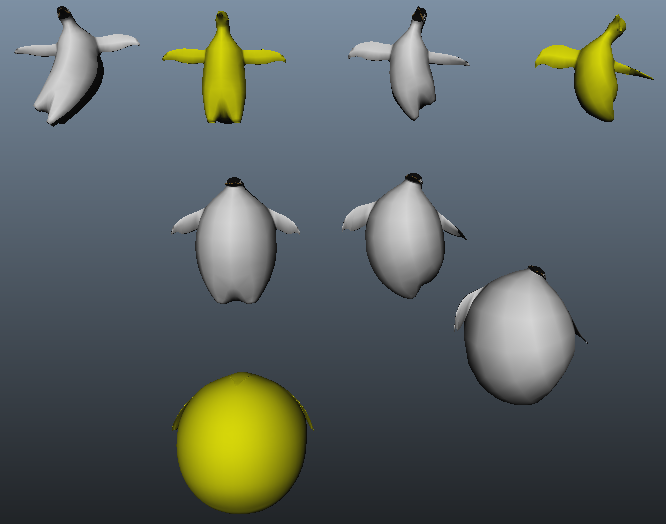}
 \caption{Three shapes (yellow) are blended to produce variations (white).
 Not only interpolation but also extrapolation (with weights $>1$ or $<0$) is possible. 
 The top-left shape is obtained by extrapolating the two yellow shapes in the top row. }
 \label{fig:penguins}
 \end{center}
\end{figure}

We also discuss an improvement of the ARAP (\S \ref{sec:arap}) in how to interpolate local transformations
(\S \ref{sec:blend}) and how to stitch fragmented tetrahedra 
by a new error function (\S \ref{sec:error}).
We demonstrate our improvement by a shape blending application (Fig. \ref{fig:penguins}).
Given an arbitrary number of isomorphic surfaces, our algorithm produces inter/extrapolation 
of the shapes according to the weights given by the user.
Roughly speaking, we define a ``linear combination'' of shapes
\[
 w_1 Q_1 + w_2 Q_2 + \cdots + w_m Q_m,
\]
where $w_i\in \R$ are weights and $Q_i$ are shapes.
In particular, when the number of shapes is two, $w_1 Q_1 + (1-w_1) Q_2$
for $0\le w_1 \le 1$ gives a morphing between them.
Note that our algorithm is highly non-linear although
we described the procedure as taking the linear combination of shapes.
We implemented the algorithm as the Autodesk Maya plugin.
Its MIT licensed source code is available at \cite{code}.

\section{Notation}
We begin with listing some notation.
We assume all tranformations are represented by real matrices, 
acting on real column vectors by the multiplication from the left.
\begin{itemize}
\item $\SO(3)$: the group of 3D rotations. Its element is a $3\times 3$ special orthogonal matrix.
\item $\Sym^+(3)$: the set of 3D shears. Its element is a $3\times 3$ positive definite symmetric matrix.
\item $\GL(3)$: the group of (invertible) 3D linear transformations consisting of compositions of rotation, shear, and reflection. 
Its element is a $3\times 3$ regular matrix. 
\item $\Aff(3)$: the group of (invertible) 3D affine transformations consisting of
compositions of rotation, shear, reflection, and translation.
Its element is a $4\times 4$ regular homogeneous matrix.
\item $\GL^+(3), \Aff^+(3)$: the subgroups of the reflection free (positive determinant) elements in the corresponding groups.
\item $\hat{A}\in \GL(3)$: the linear part ($3\times 3$ upper-left corner) of $A\in \Aff(3)$.
\item $A^t$: the transpose of a matrix $A$.
\item $|A|^2_F=\mathrm{tr}(A^t A)$: the squared Frobenius norm of a matrix $A$.
\item $\# U$:  the cardinality of a set $U$.
\end{itemize}

\section{As-rigid-as-possible deformation framework}\label{sec:arap}
In this section, we recall the ARAP framework
by describing an algorithm for shape blending.
Note that although we discuss shape blending as the primary application, 
the framework and our improvement is not limited to it.
Indeed, after being introduced in \cite{ARAP} initially as a morphing algorithm, 
the ARAP technique has been serving as one of the fundamental frameworks for
 various kinds of shape deformation applications
(see, for example, \cite{BS,probe,Sorkine,Sumner2,Sumner}).

Our problem setting is as follows.
We are given a rest shape $V_0$
and $m$ its deformations $V_j \  (1\le j\le m)$.
That is, a vertex correspondence between $V_0$ and each of $V_j \  (1\le j\le m)$ is assumed.
We would like to compute the deformation $V(w_1,\ldots,w_m)$ by blending the given shapes $\{V_j\}$ 
according to the user specified weights $\{w_j\in \R \mid 1\le j\le m\}$.
We insist that it interpolates the given shapes, i.e., 
$V(0,\ldots,0)=V_0$, and $V(w_1,\ldots,w_m)=V_k$ when $w_j=\begin{cases} 1 & (j=k) \\ 0 & (j\neq k) \end{cases}$.
Notice we allow negative weights and weights greater than one 
so that the system can not only interpolate but also extrapolate.

\begin{remark}
A basic shape blending is achieved by simply 
taking the linear combination of the coordinates of the vertices.
This method is very fast and widely used to produce
variations of shapes, in particular, facial expressions.
However, since the geometry of shapes is disregarded,
it does not always produce plausible outputs (Fig. \ref{fig:linear-blendshape}).
\begin{figure}[htbp]
 \begin{center}
   \includegraphics[height=1.8cm,keepaspectratio=true]{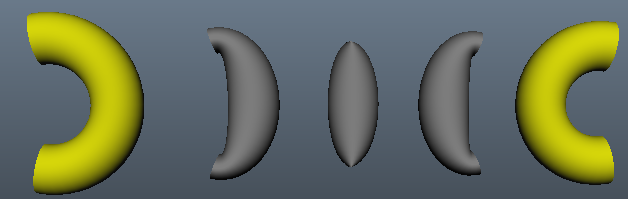}
   \includegraphics[height=1.8cm,keepaspectratio=true]{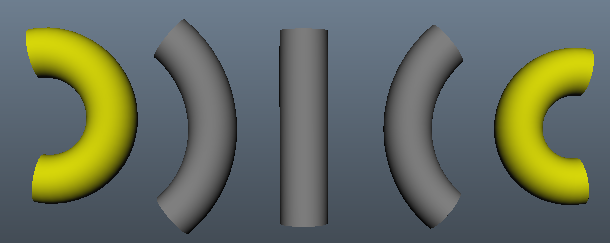}
 \caption{Interpolation between yellow shapes. Left: linear method, Right: our method}
 \label{fig:linear-blendshape}
 \end{center}
\end{figure}
The ARAP based method which we will describe below 
takes geometry into account to obtain better results.
\end{remark}

We assume that the rest shape is equipped with a non-degenerate {\em tetrahedral structure} $(V_0,\T)$.
We will discuss in \S \ref{sec:tetrisation} a method to 
associate one to a triangular mesh.
\begin{definition}
A tetrahedral structure is a pair $(V,\T)$, where 
 the vertex set $V$ consists of three dimensional vectors 
 and the set of tetrahedra
 $\T=\{T_i\mid 1\le i\le n\}$ consists of ordered tuples of four distinct vertices $T_i=(v_{i_1},v_{i_2},v_{i_3},v_{i_4})$.
Each vertex in $V$ must be contained in at least one tetrahedron.
A tetrahedral structure is said to be non-degenerate when the vertices of each tetrahedron are not co-planar.
\end{definition}
We emphasise that a triangle can be shared by three or more tetrahedra, and for this reason,
we use the terminology ``tetrahedral structure'' rather than tetrahedral mesh.

The information of a tetrahedral structure $(V,\T)$ can be packed into a collection of $4\times 4$-matrices:
\begin{equation}\label{eq:tet-matrix}
\left\{ P_i := \begin{pmatrix}
  v_{i_1}(x) & v_{i_2}(x) & v_{i_3}(x) & v_{i_4}(x) \\ 
  v_{i_1}(y) & v_{i_2}(y) & v_{i_3}(y) & v_{i_4}(y) \\ 
 v_{i_1}(z) & v_{i_2}(z) & v_{i_3}(z) & v_{i_4}(z) \\ 
 1 &1&1&1
\end{pmatrix} \Bigg\vert 1\le i \le n \right\},
\end{equation}
where $(v_{i_j}(x), v_{i_j}(y), v_{i_j}(z))^t \in \R^3$ is the vector representing the position of the vertex $v_{i_j}\in V$.

We denote by $\{ P_{0i} \mid 1\le i \le n\}$ the matrices associated to the rest shape $(V_0,\T)$.
Since $(V_0,\T)$ is assumed to be non-degenerate, all the $P_{0i}$ are regular.
For each deformation $V_j$, we use the same set of tetrahedra $\T$ to obtain $\{ P_{ji} \mid 1\le i\le n\}$.
Note that $P_{ji}$ need not be regular.
We define a series of affine transformations
\begin{equation}\label{eq:local-trans}
A_{ji} := P_{ji} P^{-1}_{0i} \ (1\le i\le n)
\end{equation}
which maps the vertices $V_0$ of the rest shape to the ones $V_j$ in the deformed shape.
Obviously, $A_{ji}v=A_{ji'}v$ when $v\in V_0$ is contained in two tetrahedra $T_i$ and $T_{i'}$.
Thus, $\{ P_{ji} \mid 1\le i\le n\}$ can be considered as a {\em piecewise linear map} defined on $(V_0,\T)$
with $(V_j,\T)$ as its image.

Now, we have $m$ piecewise linear maps $\left\{ P_{ji} \mid 1\le i\le n \right\}_{j=1}^m$
and the problem is rephrased as to blend them according to the user specified weights $w_j \ (1\le j\le m)$.
We first consider locally and blend $\hat{P}_{ji} \ (1\le j\le m)$
 for a single tetrahedron $T_i$ to obtain $C_i \in \GL^+(3)$.
Intuitively, $C_i$ stipulates the local transformation for the tetrahedron $T_i$.
We discuss a method to compute $C_i$ in \S \ref{sec:blend}.
The last step is to find a global piecewise linear map on $(V_0,\T)$, whose image we take as the output.
Since we cannot assume $C_i v$ agrees with $C_{i'}v$ for a vertex $v\in V_0$ which is contained in two tetrahedra $T_i$ and $T_{i'}$,
we have to ``stitch'' them.
What we do is to find a piecewise linear map which is closest to the collection $\{ C_i\mid 1\le i\le n\} $ with respect to an
 {\em error function}.
We discuss different error functions in \S \ref{sec:error}.
The deformed shape $V(w_1,\ldots,w_m)$ is computed as the minimiser of the error function.

In the following sections, we discuss each step in detail.

\section{Tetrisation}\label{sec:tetrisation}
In computer graphics systems, shapes are usually represented by surface meshes.
To apply the ARAP technique described in the previous section,
 we have to have a tetrahedral structure.
Here, we consider a method to build a tetrahedral structure from a given triangular surface mesh.

\begin{definition}\label{def:triangular-mesh}
For a triangular mesh, 
we denote an element of the vertex set $V$
by a three dimensional vector
and an element of the set of (face) triangles $\F$
by an ordered tuple of three vertices $(v_1,v_2,v_3)$.
For $(v_1,v_2,v_3)\in \F$, we call the ordered tuples $v_1v_2, v_2v_3$ and $v_3v_1$ the {\em oriented edges}.
A triangular mesh is said to be {\em non-degenerate} when
the vertices of each triangle are not co-linear.
\end{definition}

Given a triangular mesh, 
we would like to associate a tetrahedral structure which we can apply the ARAP framework to.
\begin{definition}
Given a non-degenerate triangular mesh $(V,\F)$.
A {\em tetrisation} of $(V,\F)$ is a tetrahedral structure
which consists of the vertex set $\bar{V}$ and the set of tetrahedra $\T$.
We require $(\bar{V},\T)$ to satisfy the following conditions:
\begin{enumerate}
\item $V\subset \bar{V}$. That is, $\bar{V}$ is obtained by adding {\em ghost vertices} to $V$.
\item Each triangle in $\F$ has to be contained in at least one tetrahedron in $\T$.
\item Each tetrahedron is non-degenerate, that is, the four vertices are not co-planar.
\end{enumerate}
\end{definition}
These conditions are exactly what are required in the ARAP framework.

We give three methods to produce tetrisation in the following.
Recall that 
the unit normal vector $n(F)$ of a triangle $F=(v_1,v_2,v_3)$ is computed by 
$\dfrac{(v_2-v_1)\times (v_3-v_1)}{|(v_2-v_1)\times (v_3-v_1)|}$,
where the denominator $|(v_2-v_1)\times (v_3-v_1)|$ is twice the area $2\text{Area}(F)$ of $F$.

\subsection{Face-normal tetrisation}\label{sec:face}
We begin with a simple method which has been commonly used in various applications.
For each triangle $F=(v_1,v_2,v_3)$ in $\F$, add the ghost vertex
\[
 v_0=\dfrac{(v_1 + v_2 + v_3)}{3} + \dfrac{(v_2-v_1)\times (v_3-v_1)}{\sqrt{|(v_2-v_1)\times (v_3-v_1)|}}
\]
and form a tetrahedron $(v_0,v_1,v_2,v_3)$.
The resulting tetrahedral structure has $\# \T=\# \F$ and $\# \bar{V} = \# V + \# \T$.

\begin{figure}[htbp]
 \begin{center}
 \includegraphics[height=2.2cm,keepaspectratio=true]{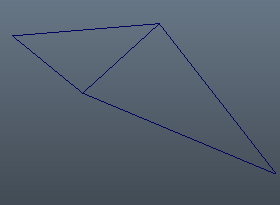}
 \includegraphics[height=2.2cm,keepaspectratio=true]{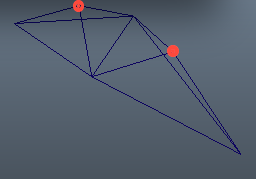}
 \caption{Left: the original surface, Right: its face-normal tetrisation. Ghost vertices are marked with a red circle.}
 \label{fig:face-normal}
 \end{center}
\end{figure}

A problem with this tetrisation when applied to the ARAP framework is that 
this does not capture the relation between adjacent triangles.
For example, consider two triangles sharing an edge as in Fig. \ref{fig:face-normal}.
Any rotation invariant error function (see \S \ref{sec:error}) with $C_1=C_2=Id$
will be minimised regardless of the angle between the two triangles.
In other words, folds do not cause any penalty in the error function.

\subsection{Edge-normal tetrisation}
We assume each oriented edge appears only once among all the triangles.
In other words, an unoriented edge should be contained at most two triangles with opposite orientations.
Also, we assume all the triangles have at least one shared edge, that is, there is no ``lone'' triangle.
(We can remove this assumption by adding ghost vertices not only for shared edges but also for all edges.
However, this is inefficient and makes no sense.)

For each shared edge $v_1v_2$, denote by $F_1=(v_1,v_2,v_3)$ and $F_2=(v_1,v_4,v_2)$ the two triangles adjacent to it.
Add a ghost vertex
\[
v_0 = \dfrac{v_1+v_2}{2} + |v_1-v_2|\dfrac{n(F_1)+n(F_2)}{|n(F_1)+n(F_2)|}
\]
and form two tetrahedra $(v_0,v_1,v_2,v_3)$ and $(v_0,v_1,v_4,v_2)$.
The resulting tetrahedral structure has $\# \T=2\cdot\#\text{(shared edges)}$ and 
$\# \bar{V} = \# V + \#\text{(shared edges)}$.
The idea of this tetrisation is to encode the angle between adjacent triangles,
which is neglected by the face-normal tetrisation.

\begin{figure}[htbp]
 \begin{center}
 \includegraphics[height=2.2cm,keepaspectratio=true]{tetrisation0}
 \includegraphics[height=2.2cm,keepaspectratio=true]{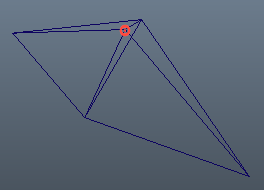}
 \caption{Left: the original surface, Right: its edge-normal tetrisation. Ghost vertex is marked with a red circle.}
 \end{center}
\end{figure}

\subsection{Vertex-normal tetrisation}
We assume that every vertex has a neighbourhood homeomorphic to the plane or the half plane.
In other words, the mesh is a manifold (with boundary).
Also, we assume all the triangles have at least one shared vertex.
(Again, we can remove this assumption as in the previous subsection.)

For each shared vertex $v$, denote the adjacent triangles by $F_1,F_2,\ldots,$ and $F_k$.
Add a ghost vertex
\[
 v_0=v + \sqrt{\sum_{i=1}^k \text{Area}(F_i)} \dfrac{n(F_1)+\cdots +n(F_k)}{|n(F_1)+\cdots +n(F_k)|}
\]
and form $k$ tetrahedra by adding $v_0$ to the triangles $F_i \ (1\le i\le k)$.
The resulting tetrahedral structure has $\# \T=3\# \F-\#\text{(non-shared vertices)}$ and 
$\# \bar{V} = \# V + \#\text{(shared vertices)}$.
An advantage of this method is that it extends straightforwardly to 
general polyhedral meshes.
The idea of this tetrisation is to encode the angle around internal vertices,
which is neglected by the face-normal tetrisation.
\begin{figure}[htbp]
 \begin{center}
 \includegraphics[height=2.0cm,keepaspectratio=true]{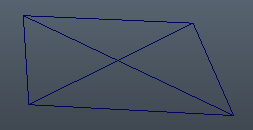}
 \includegraphics[height=2.0cm,keepaspectratio=true]{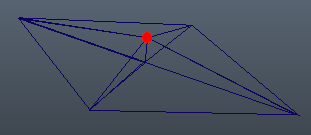}
 \caption{Left: the original surface, Right: its vertex-normal tetrisation. Ghost vertex is marked with a red circle.
 Ghost vertices on the boundary are omitted for simplicity.}
 \end{center}
\end{figure}

\section{Blending linear maps}\label{sec:blend}
In this section, we discuss how to blend local transformations 
$\hat{A}_{1i},\hat{A}_{2i}, \ldots, \hat{A}_{mi}\in \GL^+(3)$
with regard to the weights $w_1,\ldots,w_m\in \R$ to obtain the blended local transformation
$C_i\in \GL^+(3)$.
For this purpose, we use 
a function $\Blend: \R^m\times (\GL^+(3))^m \to \GL^+(3)$ 
which satisfies the obvious requirement for interpolation. Then, we set
\[
 C_i:=\Blend( w_1,\ldots,w_m, \hat{A}_{1i}, \hat{A}_{2i}, \ldots, \hat{A}_{mi} ).
\]
We investigate two such interpolation functions.

First, decompose each $\hat{A}_{ki}$ by the polar decomposition (see, for example, \cite{Higham,polar})
\[
\hat{A}_{ki} = R_{ki}S_{ki}
\]
where $R_{ki}\in \SO(3)$ is the rotation and
$S_{ki} \in \Sym^+(3)$ is the shear.
In \cite{Sumner}, they suggest
\[
\Blend_P( w_1,\ldots,w_m, \hat{A}_{1i}, \ldots, \hat{A}_{mi} )
= \exp(\sum_{k=1}^m w_k\log(R_{ki})) \left( \sum_{k=1}^m w_k S_{ki} + \left(1-\sum_{k=1}^m w_k \right) I \right),
\]
where $\log$ is the principal matrix logarithm and $I$ is the identity matrix\footnote{
The term involving $I$ is for normalisation and it enforces
$\Blend_P( 0,\ldots,0, \hat{A}_{1i}, \hat{A}_{2i}, \ldots, \hat{A}_{mi} )=I$.}. 
This coincides with the one used in \cite{ARAP} when $m=1$.
On the other hand, we suggest
\begin{equation}\label{eq:interpolation}
\Blend_C( w_1,\ldots,w_m, \hat{A}_{1i},  \ldots, \hat{A}_{mi} )
= \exp(\sum_{k=1}^m w_k\log^c(R_{ki}))  \exp(\sum_{k=1}^m w_k\log(S_{ki})),
\end{equation}
where $\log^c$ is the ``continuous'' logarithm such that it chooses the nearest 
branch of logarithm to the adjacent tetrahedra when $i$ varies (see \cite{Kaji-Ochiai} for details).
The indeterminacy of $\log$ for $\SO(3)$ is in the rotation angle
and $\log^c$ chooses the angle continuously for adjacent tetrahedra.
Note that \cite{Kaji-Ochiai} provides a direct and fast formula for $\Blend_C$ which 
does not require the polar decomposition.

\begin{figure}[htbp]
 \begin{center}
   \includegraphics[height=2.2cm,keepaspectratio=true]{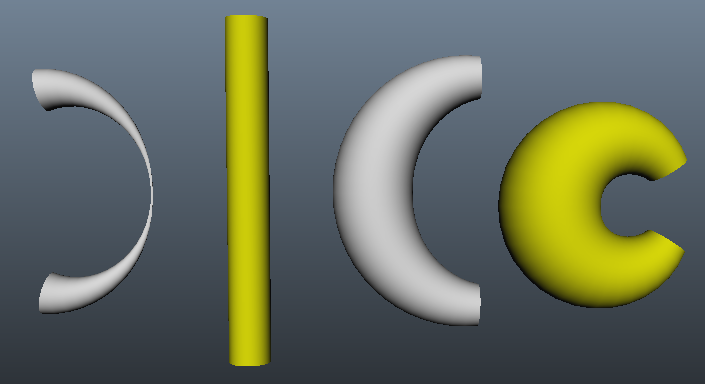}
   \includegraphics[height=2.2cm,keepaspectratio=true]{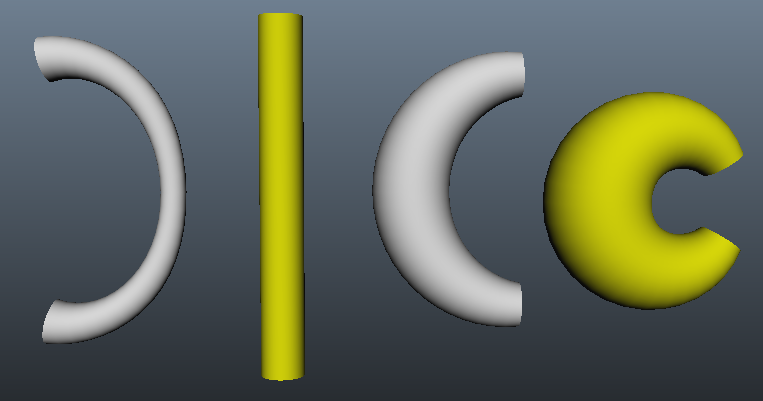}
 \caption{Interpolation/extrapolation of yellow shapes.
 Left: with $\Blend_P$ function in \cite{Sumner}, the extrapolated shape on the left is degenerate.
 Right: with our $\Blend_C$ function, the extrapolated shape is non-degenerate.
 }
 \label{fig:sym-linear}
 \end{center}
\end{figure}

\begin{figure}[htbp]
 \begin{center}
   \includegraphics[height=2.2cm,keepaspectratio=true]{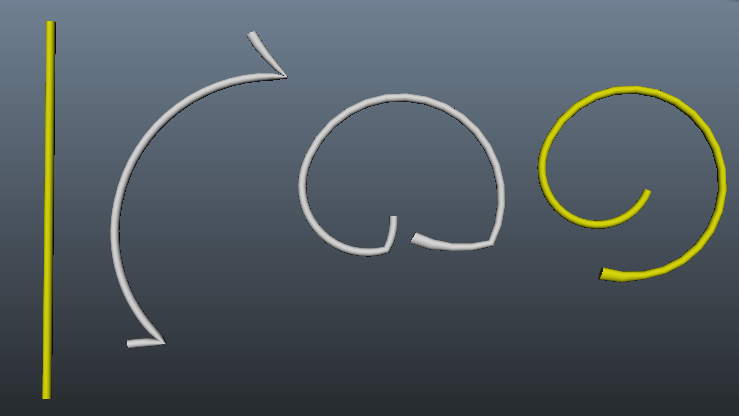}
   \includegraphics[height=2.2cm,keepaspectratio=true]{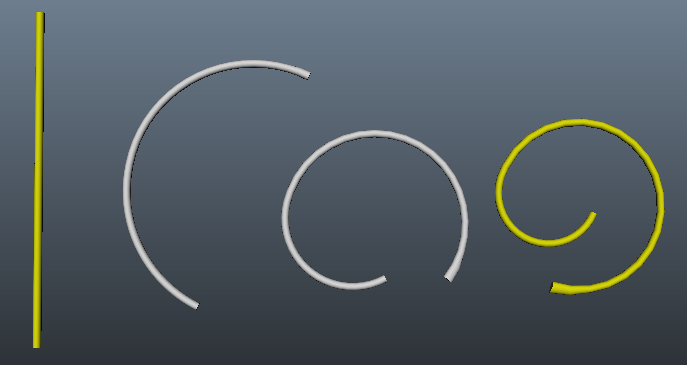}
 \caption{Interpolation of yellow shapes.
 Left: with $\Blend_P$ function in \cite{Sumner}, some parts try to rotate inconsistently. 
 Right: with our $\Blend_C$ function,
 local rotations are appropriately handled to produce a smooth interpolation}
 \label{fig:large-rotation}
 \end{center}
\end{figure}

They look similar but there are two significant differences;
blending for the shear part and logarithm for $\SO(3)$.
The value of $\Blend_P$ can fall out of $\GL^+(3)$ due to the linear blending of the shear part,
which causes distortion in the output (Fig. \ref{fig:sym-linear}).
The use of the continuous logarithm enables the system to produce a smoother morph among shapes which
performs large rotation in between (Fig. \ref{fig:large-rotation}).
Note that in \cite{ARAP} which discusses morphing of two shapes,
they suggest to use the quaternions and SLERP (\cite{SLERP}) to interpolate the rotation part
and the linear interpolation for the shear part.
(With three or more shapes, one can use the linear blending of the quaternions for the rotation part as in \cite{DLB}.)
However, this method shows similar deficiency as $\Blend_P$.

\section{Error function}\label{sec:error}
In this section, we consider error functions to stitch fragmented tetrahedra.
Fix the vertex positions $V_0$ of the rest shape and the local transformations $\{C_i\mid 1\le i\le n\}$
of the tetrahedra.
An error function is a function of the deformed vertex positions $V'$.
By Equation \eqref{eq:local-trans}, a piecewise linear map $\{A_i \mid 1\le i\le n\}$ and $V'$
 are linearly related and we identify them.
In \cite{ARAP}, they introduced
\begin{equation}\label{translation-invariant-error}
 E_T(V',\{C_i\})=\sum_{i=1}^n |\hat{A}_i - C_i|^2_F
\end{equation}
and it has been used in many of the ARAP based shape deformation applications including \cite{nway,Sumner2,Sumner,Yu}.
Note that the function is translation invariant but not rotation invariant.
Rotation invariance is sometimes preferable in shape deformation (see, for example, \cite{Lipman,Sorkine} and Fig. \ref{fig:comparison}).
We propose an alternative error function which is rotation and translation invariant:
\begin{equation}\label{shear-error}
 E_S(V',\{C_i\}) = \sum_{i=1}^n |S(\hat{A}_i) - S(C_i)|^2_F,
\end{equation}
where $S(X)$ for $X\in \GL(3)$ is the shear factor of the polar decomposition of $X$
(see \cite{polar}).
Intuitively, this error function measures how much each tetrahedron is distorted.
Despite the simplicity and its invariance property, 
$E_S$ has not been considered in the literature as far as the author is aware.
We believe this error function gives a good alternative to $E_T$ in some applications
(see Fig. \ref{fig:comparison}).

\begin{remark}
We can assign a weight $W_i\in \R$ to each tetrahedron $T_i$
to specify its contribution to the error function.
It is done simply by replacing the summation $\sum_{i=1}^n$ with the weighted one
$\sum_{i=1}^n W_i$ in the definitions of the error functions.
For notational simplicity, we omit them in this paper.
\end{remark}

As we described in \S \ref{sec:arap}, we define the output as the minimiser of the error function.
In other words, we compute the piecewise linear function $\{A_i \mid 1\le i\le n\}$ which is closest to 
$\{C_i\mid 1\le i\le n\}$ with respect to the error function.
Computing the minimiser for $E_T$ is reduced to solving a sparse linear system (see \cite{ARAP,Sumner}).
For $E_S$, the computation is not linear. An iterative way similar to \cite{Sorkine} is given as follows:
%
\begin{enumerate}
\item Compute the minimiser of $E_T(V',\{C_i\})$ and set $\hat{A}_i$.
\item Compute the polar decomposition $\hat{A}_i=R_i S_i$.
\item Compute the minimiser of $E_T(V',\{R_i S(C_i)\})$ to update $\{ \hat{A}_i \}$.
\item Repeat (2) and (3) until $\{ \hat{A}_i \}$ converge.
\end{enumerate}

Note that there is some indeterminacy of the minimiser coming from the symmetry of the error function.
For example, any translation of a minimiser is also a minimiser.
To obtain a unique minimiser, one can impose additional constraints; for $E_T$
fixing the position of the barycentre
and for $E_S$
fixing the position of the barycentre and the orientation of some tetrahedra.

\section{Implementation}\label{N-way}
\begin{figure}[htbp]
 \begin{center}
   \includegraphics[height=4.0cm,keepaspectratio=true]{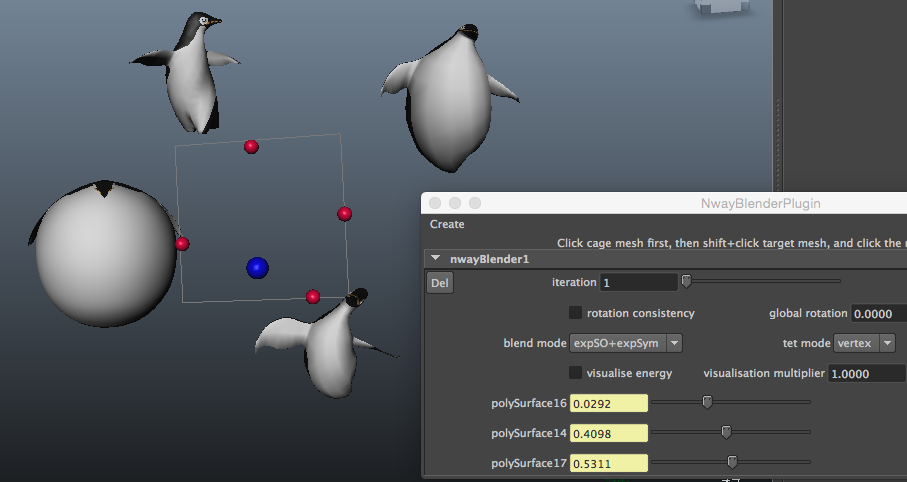}
 \caption{Our Maya plugin}\label{fig:maya}
 \end{center}
\end{figure}

We implemented our algorithm as the Autodesk Maya plugin (\cite{code}).
In our system, the user can specify the weight for each shape with sliders,
or the ball controller which computes the weights by \cite{Floater} from the configuration of the balls
representing the shapes (Fig. \ref{fig:maya}).

\begin{figure}[htbp]
 \begin{center}
   \includegraphics[height=2.2cm,keepaspectratio=true]{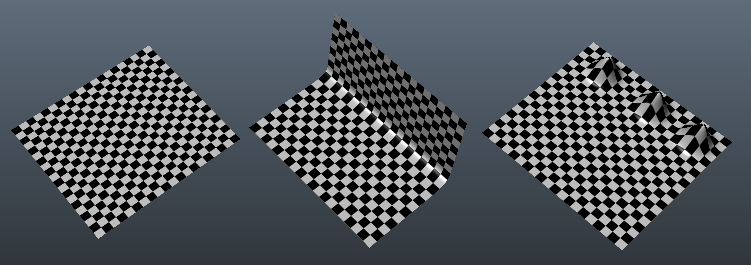} \\
   \includegraphics[height=2.6cm,keepaspectratio=true]{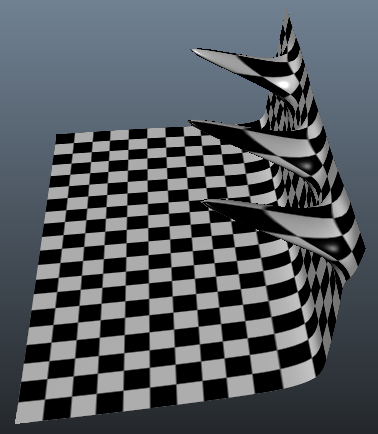}
   \includegraphics[height=2.6cm,keepaspectratio=true]{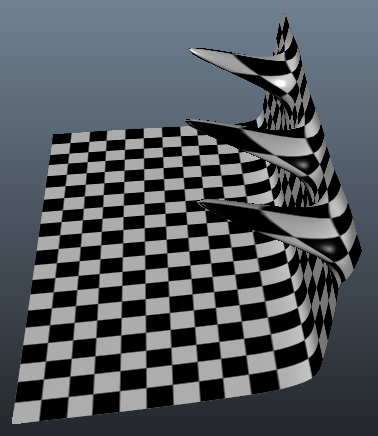}
   \includegraphics[height=2.6cm,keepaspectratio=true]{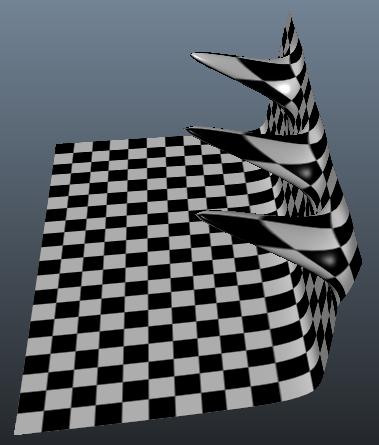} \\
   \includegraphics[height=2.6cm,keepaspectratio=true]{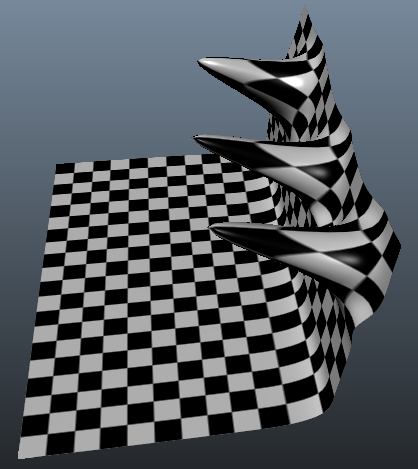}
   \includegraphics[height=2.6cm,keepaspectratio=true]{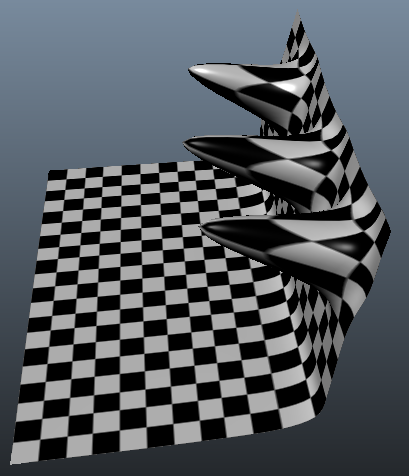}
   \includegraphics[height=2.6cm,keepaspectratio=true]{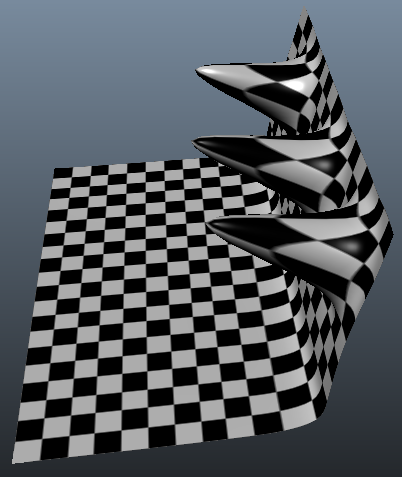}
 \caption{Top row from left to right: rest shape $V_0$ and its two deformations $V_1$ and $V_2$ to be blended with weights $w_1=1.0$ and $w_2=1.5$.
 Second row from left to right: results obtained by face-normal, edge-normal, and vertex-normal tetrisation with $E_T$, 
 Third row: same as the second row but with $E_S$.}
 \label{fig:comparison}
\end{center}
\end{figure}

The ARAP framework was also applied to shape blending in  
\cite{nway} in the 2D setting and 
in \cite{Sumner} in the 3D setting.
We demonstrate our improvement discussed in \S \ref{sec:tetrisation} and \S \ref{sec:error}
by comparing with \cite{Sumner}.
First, we note that in \cite{Sumner},
(i) the face-normal tetrisation,
(ii) the error function $E_T$,
(iii) and the blending function $\Blend_P$
are used.
We have already seen the difference between the blending functions $\Blend_P$ and $\Blend_C$ 
at the end of \S \ref{sec:blend}.
We will turn our attention to (i) and (ii).
Fig. \ref{fig:comparison} visually compares different tetrisations
in \S \ref{sec:tetrisation}
and the error functions $E_T$ and
 $E_S$ in \S \ref{sec:error}.
We observe that $E_S$ produces more natural results than $E_T$ but much slower as we see in Table \ref{table:timing}.
With $E_S$, the face-normal tetrisation causes extra wrinkles compared to the edge-normal
 and the vertex-normal tetrisations.
 As far as we experimented, it depends on the character of shapes to be blended 
 which tetrisation gives the best result.
 In general, with $E_T$ the output is more or less similar regardless of the choice of tetrisation. 
 With $E_S$, the vertex-normal tetrisation seems to be a good choice.

Table \ref{table:timing} shows a timing comparison for different tetrisations and error functions.
We blended two 3D models each with 26k triangles
 on a Macbook Air with 1.7GHz Intel Core i7 and 8GB memory.
 Initialisation part involves the Cholesky decomposition of the space matrix necessary to solve the minimiser of the 
 error functions. This is computed only once in the initialisation process.
 Note that the matrix is dependent on the tetrahedral structure but 
 independent of the choice of the error function.
 Runtime part consists of finding the minimiser of the error functions and the computation of $\Blend$ functions.

\begin{table}[htbp]
\caption{Timing comparison}
\label{table:timing}
\begin{tabular}{c|ccc|ccc}
\hline\noalign{\smallskip}
 & face $E_T$ & edge $E_T$ & vertex $E_T$ & 
  face $E_S$ & edge $E_S$ & vertex $E_S$  \\
\noalign{\smallskip}\hline\noalign{\smallskip}
Initialisation (sec) & 0.1976  & 0.3080  & 0.3556  & 0.1976  & 0.3080  & 0.3556 \\
Runtime with $\Blend_P$ & 45.45 fps & 27.43 fps & 30.86 fps & 11.66 fps & 4.132 fps & 4.762 fps\\
Runtime with $\Blend_C$ & 48.46 fps & 29.31 fps & 31.46 fps & 12.19 fps & 4.576 fps & 5.037 fps\\
\noalign{\smallskip}\hline\noalign{\smallskip}
\end{tabular}
\end{table}

\subsection*{Acknowledgement}
This work was partially supported by the Core Research for Evolutional Science and Technology (CREST) Program 
titled ``Mathematics for Computer Graphics'' of the Japan Science and Technology Agency (JST), 
by KAKENHI Grant-in-Aid for Young
     Scientists (B) 26800043, and by JSPS Postdoctoral Fellowships for Research Abroad.

%


\begin{thebibliography}{99.}

\bibitem{ARAP} 
M. Alexa, D. Cohen-Or, and D. Levin,
\textit{As-Rigid-As-Possible Shape Interpolation},
  Proc. ACM SIGGRAPH 2000, pp. 157--164 (2000)

\bibitem{nway}
W. Baxter, P. Barla, and K. Anjyo,
\textit{N-way morphing for 2D animation}, 
Computer Animation and Virtual Worlds 20, 2-3, 79--87  (2009).


\bibitem{BS}
M. Botsch and O. Sorkine,
\textit{On Linear Variational Surface Deformation Methods},
IEEE Transactions on Visualization and Computer Graphics 14(1),
213--230 (2008)

\bibitem{Floater}
M. S. Floater,
\textit{Mean value coordinates},
 Computer Aided Geometric Design 20(1), 19--27 (2003)

%

\bibitem{Higham}
N. Higham,
{\it Computing the Polar Decomposition--With Applications},
 SIAM J. Sci. and Stat. Comp. 7(4), 1160--1174 (1986)



\bibitem{code}
S. Kaji, 
\textit{An N-way morphing plugin for Autodesk Maya}, 
\verb+https://github.com/shizuo-kaji/NWayBlenderMaya+

\bibitem{probe}
S. Kaji and G. Liu, 
\textit{Probe-type deformers}, 
Mathematical Progress in Expressive Image Synthesis II, 
Springer-Japan, 63--77 (2015)

\bibitem{Kaji-Ochiai}
S. Kaji and H. Ochiai, 
\textit{A concise parametrisation of affine transformation}, 
preprint, arXiv:1507.05290

\bibitem{DLB} 
L. Kavan, S. Collins, J. \v{Z}\'{a}ra, and C. O'Sullivan,
\textit{Geometric Skinning with Approximate Dual Quaternion Blending}.     
ACM Transaction on Graphics 27(4), 105:1--105:23 (2008)


\bibitem{Lipman}
Y. Lipman, O. Sorkine, D. Levin, and D. Cohen-Or.,
{\it Linear rotation-invariant coordinates for meshes},
ACM Transaction on Graphics 24(3), 479--487 (2005)

\bibitem{SLERP} K. Shoemake,
\textit{Animating rotation with quaternion curves}, ACM SIGGRAPH, pp. 245--254, 1985.

\bibitem{polar} K. Shoemake and T. Duff,
{\it Matrix animation and polar decomposition},
Proc. Graphics interface '92, Kellogg S. Booth and Alain Fournier (Eds.). Morgan Kaufmann Publishers Inc., San Francisco, CA, USA, 258--264 (1992)

\bibitem{tetgen}
H. Si,
\textit{TetGen, a Delaunay-based quality tetrahedral mesh generator},
 ACM Trans. Math. Softw. 41(2), 11:1--11:36 (2015)

\bibitem{Sorkine}
O. Sorkine and M. Alexa,
{\it As-rigid-as-possible surface modeling},
Proc. Eurographics SGP '07,
Eurographics Association, Aire-la-Ville, Switzerland, Switzerland, 109--116 (2007)


\bibitem{Sumner2}
R. W. Sumner and J. Popovi\'{c},
\textit{Deformation transfer for triangle meshes},
ACM Transactions on Graphics 23(3), 399--405 (2004)

\bibitem{Sumner}
R. W. Sumner, M. Zwicker, C. Gotsman, and J. Popovi\'{c},
\textit{Mesh-based inverse kinematics},
ACM Transaction on Graphics 24(3), 488--495 (2005)


\bibitem{Yu}
Y. Yu, K. Zhou, D. Xu, X. Shi, H. Bao, B. Guo, and H-Y Shum, 
{\it Mesh Editing with Poisson-based Gradient Field Manipulation},
ACM Transaction on Graphics 23(3), 644--651 (2004)

\end{thebibliography}
\end{document}